\documentclass[10pt,conference]{IEEEtran}
\IEEEoverridecommandlockouts

\usepackage{arydshln}
\usepackage{makecell}
\usepackage{pifont}
\usepackage{cite}
\usepackage{amsmath,amssymb,amsfonts}
\usepackage{algorithm}
\usepackage{algorithmic}
\usepackage{graphicx}
\usepackage{textcomp}
\usepackage{xcolor}
\usepackage{xspace}
\usepackage{booktabs}
\usepackage{listings}
\usepackage{url}
\usepackage{multirow}

\newcommand{\ourapproach}{\textsc{PRAXIS}\xspace}

\lstdefinestyle{py}{%
  language=Python, basicstyle=\scriptsize\ttfamily,
  keywordstyle=\color{blue!55!black}, commentstyle=\color{green!40!black}\itshape,
  stringstyle=\color{orange!75!black}, showstringspaces=false, breaklines=true,
  frame=single, framerule=0.3pt, captionpos=b, xleftmargin=3pt, aboveskip=4pt, belowskip=2pt}
\graphicspath{{figures/}}

\newenvironment{rqlist}{%
  \begin{list}{\textbullet}{%
    \setlength{\topsep}{2pt}\setlength{\partopsep}{0pt}%
    \setlength{\parsep}{0pt}\setlength{\itemsep}{1pt}%
    \setlength{\leftmargin}{1.4em}\setlength{\labelwidth}{1em}%
    \setlength{\labelsep}{0.4em}}%
}{%
  \end{list}%
}

\begin{document}

\title{\ourapproach: Graph-Grounded Tacit Knowledge for Domain Coding Agent}

\author{
\IEEEauthorblockN{
Xue Jiang\textsuperscript{1},
Tianyu Zhang\textsuperscript{1},
Lingwei Wu\textsuperscript{1},
Ziyu Wang\textsuperscript{1},
Yuan Sui\textsuperscript{2},
Hao Zhu\textsuperscript{1},
Wenpin Jiao\textsuperscript{1},\\
Zhi Jin\textsuperscript{1},
Ge Li\textsuperscript{1,*},
Yihong Dong\textsuperscript{3}
}
\IEEEauthorblockA{
\textsuperscript{1}Key Laboratory of High Confidence Software Technologies (PKU), Ministry of Education;\\
School of Computer Science, Peking University\\
\textsuperscript{2}School of Computing, National University of Singapore\\
\textsuperscript{3}School of Computer Science, Shanghai Jiao Tong University
}
}

\maketitle

{\renewcommand{\thefootnote}{}\footnotetext{\textsuperscript{*}Corresponding author.}}

\begin{abstract}
LLM agents have achieved strong performance on general software engineering tasks, yet struggle with domain-specific code generation. We identify the root cause as the agent's lack of tacit knowledge, including domain-specific business rules, interface contracts, and operational conventions that developers internalize through practice but never document. This knowledge is deeply buried beneath the domain code, dispersed across code entities and their dependency relations, and invisible to the agent that lacks it. These properties make tacit knowledge inherently difficult to retrieve
or learn. In this work, we propose PRAXIS, a framework that enables agents to systematically extract, represent, and reuse tacit knowledge for domain code generation. PRAXIS acquires tacit knowledge by simulating human development workflows within the target codebase, distills it into structured units organized on the code dependency graph, and proactively surfaces it to the agent at the point of code interaction. Extensive experiments demonstrate that PRAXIS outperforms state-of-the-art agents equipped with powerful agentic search capabilities, as well as experience-based and skill-based methods. The approach integrates seamlessly into various agent frameworks and LLMs with consistent performance improvements, and supports continual evolution with performance steadily
scaling as practice accumulates.

\end{abstract}

\begin{IEEEkeywords}
LLM-based Agent, Code Generation, Domain Knowledge
\end{IEEEkeywords}

\section{Introduction}
\label{sec:intro}

LLM-based agents have made remarkable strides on general software engineering tasks, with leading systems resolving the majority of problems on widely used benchmarks~\cite{swebench, sweagent, openhands, specrover}. However, real-world software development is dominated by domain code generation~\cite{domcoder, DBLP:journals/tosem/DongJJL24, repocoder, codereval, crosscodeeval, repobench}, which requires project-specific knowledge that is absent from the model's training data. In this setting, even the strongest agents fail to maintain their general-purpose performance, with accuracy declining sharply on domain-specific datasets~\cite{kocobench, domaincodebench,
evocodebench}. Closing this gap is therefore a key open problem.

We investigate the underlying causes of this performance gap. We find that even when the agent is provided with all ground-truth code that the target implementation depends on, it still largely fails, suggesting that the true difficulty lies not in what is visible in the codebase but in what is hidden beneath it.
Manual analysis of the failure cases reveals that the root cause lies in the agent's lack of \emph{tacit knowledge}: domain-specific business rules, interface contracts, and operational conventions that developers internalize through sustained engagement with a codebase~\cite{latoza2006, siegmund2014, robillard2009}. Such knowledge dictates how code ought to be correctly written, yet it exists only in developers' minds and is rarely recorded in the repository's documentation, comments, or source code. In a further attempt, we let the agent explore the codebase and summarize the domain knowledge it discovers, but observe only marginal improvement. Our analysis indicates that tacit knowledge resists both extraction and utilization due to three inherent properties. First, it is deeply buried beneath the domain code and only surfaces during actual development workflows. Second, it is dispersed across individual code entities and their
dependency relations, with its effects propagating along dependency paths rather than being fixed in a single point.
Third, an agent that lacks tacit knowledge does not realize it is missing, and therefore will not actively seek it.
These three properties constitute the central challenges that this paper aims to address.

This paper addresses three core problems: how to extract tacit knowledge, how to represent it, and how to reuse it. For extraction, we have the agent simulate the workflow of a human developer by actually writing code within the target codebase, encountering errors, and iteratively revising~\cite{reflexion, mckeeman1998}. This process exposes knowledge that would otherwise remain deeply hidden and difficult to access through reading or summarization alone. For representation, because tacit knowledge is often distributed across individual code entities and their dependency relations, storing it as a flat list of text entries would lose its structural information. We instead anchor each piece of knowledge to the corresponding code entity and organize all entries over the code dependency graph~\cite{pdg, sdg, astgraphrag} to preserve their propagation paths. For reuse, because an agent that lacks tacit knowledge struggles to recognize its own deficit, mechanisms that rely on the agent to search knowledge will fail. We proactively deliver tacit knowledge to agent in the appropriate context, allowing knowledge to reach the agent rather than requiring the agent to seek it out.

In this paper, we propose \ourapproach, a novel framework for domain code generation. \ourapproach focuses on the unique properties of tacit knowledge in domain code and enhances the agent's ability to generate domain-specific code by systematically extracting, representing, and reusing such knowledge. In-Domain Development Practice introduces a test-time practice stage in which the agent simulates human development workflows within the target codebase to actively surface tacit knowledge. Structured Knowledge Acquisition distills the tacit knowledge captured in the agent's practice trajectories into structured knowledge units. Graph-Grounded Knowledge Organization subsequently anchors each knowledge unit to its corresponding code entity on the dependency graph, propagating them along dependency relations to preserve cross-component constraints. Tacit Knowledge Injection completes the pipeline through a proactive approach orthogonal to conventional passive retrieval, automatically surfacing relevant knowledge as the agent encounters the associated code entities. The framework further supports online continual evolution, where new knowledge discovered by the agent during real tasks is incorporated in real time, allowing its domain expertise to accumulate steadily through practice.

Extensive experimental results demonstrate that \ourapproach achieves significant advantages over existing coding agents, SOTA experience-based and skill-based methods on domain code generation, outperforming the best-performing baseline by a relative 16.7\% on Pass@1.
The approach exhibits consistent effectiveness across different agent frameworks and LLMs of varying families. Moreover, as the scale of practice increases and online knowledge evolution continues, agent performance shows a steady upward trend. We also conduct a systematic quality assessment and typological analysis of the extracted tacit knowledge, revealing its concrete forms and distribution, providing empirical evidence for understanding what tacit knowledge actually is, and validating the effectiveness of the extracted knowledge. 

Our main contributions are as follows:
\begin{itemize}
  \item We identify tacit knowledge as the fundamental bottleneck in
    domain code generation, and characterize three properties
    that distinguish it from general knowledge and make it particularly
    resistant to existing approaches.
  \item We propose \ourapproach, which systematically addresses the
    extraction, representation, and reuse of tacit knowledge through
    four stages: In-Domain Development Practice, Structured
    Knowledge Acquisition, Graph-Grounded Knowledge Organization, and
    Tacit Knowledge Injection.
  \item Extensive experiments demonstrate that \ourapproach
    outperforms existing coding agents as well as experience-based
    and skill-based methods, generalizes consistently across agent
    frameworks and LLMs, and exhibits steady
    performance scaling as practice accumulates.
\end{itemize}

As general-purpose code generation capabilities advance rapidly, domain knowledge is becoming the final barrier to deploying AI agents in real development. \ourapproach offers a solution to this critical bottleneck, taking a meaningful step toward building autonomous agents that continuously accumulate domain knowledge in real-world software engineering environments. Our source code and data are available at \url{https://github.com/jiangxxxue/PRAXIS}.
\section{Motivation}
\label{sec:motivation}

\begin{figure}[t]
  \centering
  \includegraphics[width=0.85\columnwidth]{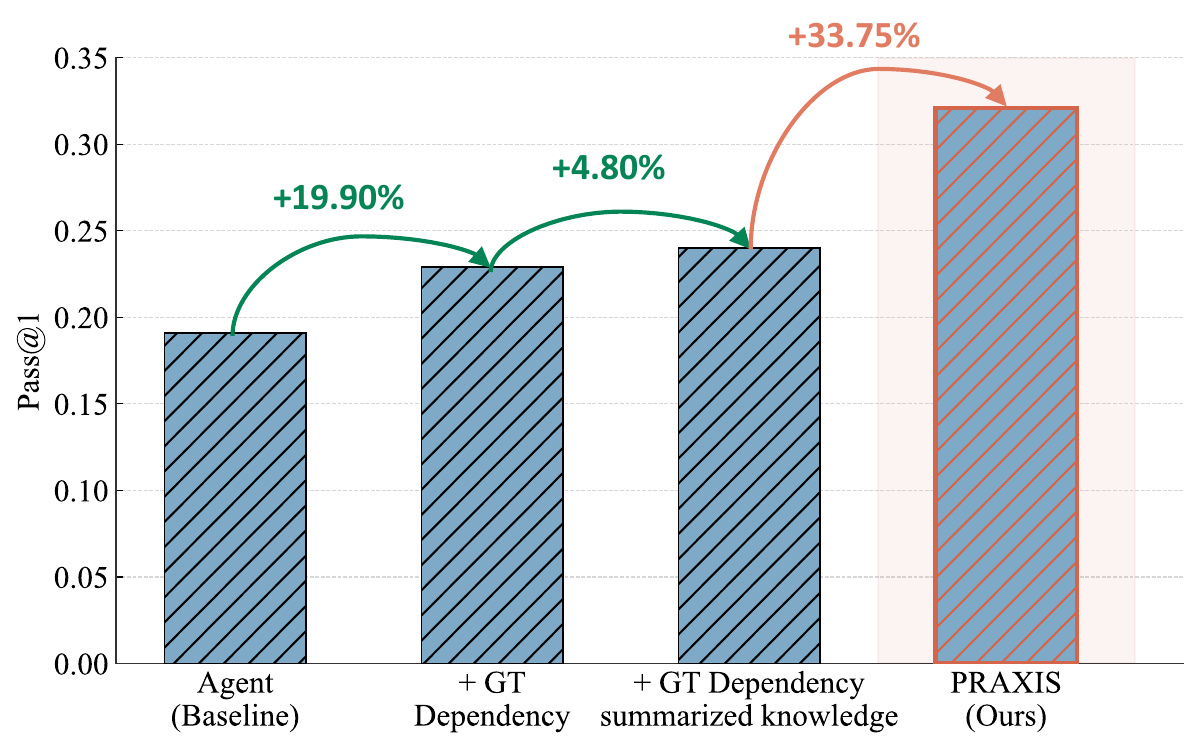}
  \caption{Results of preliminary experiments.}
  \label{fig:motivation}
\end{figure}
 
\subsection{Preliminary Experiments}
\label{sec:motivation:exp}

To explore the performance gap in domain code generation, we conduct controlled experiments on KoCo-Bench~\cite{kocobench}, a representative domain code generation benchmark, using OpenHands~\cite{openhands}, a widely adopted open-source coding agent, with DeepSeek-V3.2~\cite{deepseekv3_2} as the base model.
Under standard conditions, the agent achieves a Pass@1 of only 19.1\% (Figure~\ref{fig:motivation}). This performance gap can be attributed to two categories of factors. Internal factors refer to the model's inherent code generation capabilities, while external factors refer to the level of support the agent provides to the model during its development workflow. In this paper, we focus on the external factors. We attempt to construct a near-perfect code context for the model by directly supplying the source code of all ground-truth dependencies. This yields a 19.9\% relative improvement, confirming that better external support does help. However, the overall performance remains low. Examining the results, we find that the model frequently fails to effectively utilize these dependencies. Even when the correct dependency is directly available, the agent struggles to understand how to properly incorporate it into its own implementation. 
To investigate this further, we let the agent explore the repository and summarize the domain knowledge of these dependencies, and supply the summaries alongside the source code. This yields only a negligible 4.8\% relative improvement. This knowledge is difficult for the agent to acquire through any observable artifact in the code repository. In practice, human developers internalize such knowledge only through long-term development within the project. Furthermore, the agent is unable to identify or articulate the specific knowledge it lacks. We refer to this missing category as \emph{tacit knowledge} and analyze its concrete forms below.

\subsection{Case Analysis}
\label{sec:motivation:case}

We show a representative task on which the agent is given all ground-truth dependencies and extracted knowledge, yet produces an incorrect implementation. The task asks for \texttt{\_create\_rag\_instance}, a factory method whose job is to configure callbacks and return a fully initialized RAG engine. The agent's implementation executes without error. It builds the configuration, defines the LLM and vision callbacks following the exact patterns in the dependency source, handles the optional reranker, and returns the engine. Nevertheless, the implementation passes none of the test cases due to two unwritten contracts.

\textbf{Contract (A): Chunk augmentation before embedding.} The agent calls the embedding service directly:
\begin{lstlisting}[language=Python,basicstyle=\ttfamily\scriptsize]
# Agent's implementation
embeddings = await self.embedding_wrapper.embed(texts)
\end{lstlisting}

\noindent The ground truth augments each chunk with the project's retrieval prompt template before the call:

\begin{lstlisting}[language=Python,basicstyle=\ttfamily\scriptsize]
# Ground truth
augmented = [self.retrieval_prompt_template.format(text=t) for t in texts]
embeddings = await self.embedding_wrapper.embed(augmented)
\end{lstlisting}

\noindent The agent calls the embedding wrapper in accordance with its interface specification, and the call completes without error. However, this project's retrieval relies on a project-specific prompt template. The query side independently wraps user queries with this template before encoding. To keep the resulting vectors aligned, the indexing side must also apply the same template to chunks before embedding. There is no place in the codebase that states this mutual constraint. It is a convention embedded in the structure of the codebase, never explicitly exposed. As a result, the agent produces an implementation that appears correct but silently degrades retrieval quality.

\textbf{Contract (B): LLM callback must be registered with the project's model router.}
The agent constructs the callback and passes it directly to the engine:

\begin{lstlisting}[language=Python,basicstyle=\ttfamily\scriptsize]
# Agent's implementation
async def _create_rag_instance(self, config):
    llm_func = self._build_llm_callback(config)
    embedding_func = self._build_embedding_func(config)
    return RAGAnything(config=config,
                       llm_model_func=llm_func,
                       embedding_func=embedding_func)
\end{lstlisting}

\noindent The ground truth additionally registers the LLM callback with the project's shared model router:

\begin{lstlisting}[language=Python,basicstyle=\ttfamily\scriptsize]
# Ground truth
async def _create_rag_instance(self, config):
    llm_func = self._build_llm_callback(config)
    embedding_func = self._build_embedding_func(config)
    self.llm_router.register(llm_func, config.model_name)
    return RAGAnything(config=config,
                       llm_model_func=llm_func,
                       embedding_func=embedding_func)
\end{lstlisting}

\begin{figure}[h!]
  \centering
  \includegraphics[width=0.75\columnwidth]{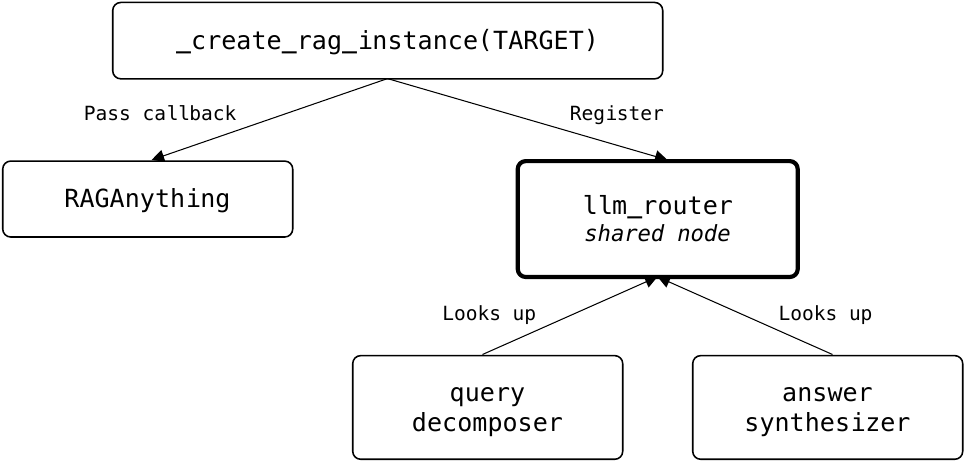}
  \caption{Dependency Relationship of Contract (B).}
  \label{fig:dep-contractB}
\end{figure}

\noindent The factory's job is to configure and return a RAG engine, and the agent's implementation fulfills this correctly. However, the agent overlooks the coordination required with other modules in the project. In this project, the RAG engine receives its LLM callback through direct parameter passing, but other pipeline stages such as the query decomposer and the answer synthesizer resolve their LLM callback at runtime by looking it up from a shared \texttt{llm\_router}. The factory must register its callback with this router so that these downstream stages can access it. This is a project-level convention for coordinating model access across independently developed pipeline stages. Without registration, these stages silently fall back to a default model, producing incorrect results. As shown in Figure~\ref{fig:dep-contractB}, although this constraint is invisible from the factory's own interface, the factory and its downstream consumers are structurally connected through the shared \texttt{llm\_router} node in the dependency graph.

\subsection{Three Properties of Tacit Knowledge}
\label{sec:motivation:properties}

Generalizing from this case and the broader set of failures in our experiments, we identify three properties of tacit knowledge that make it inherently difficult to retrieve and learn.

\textbf{Property 1: Tacit knowledge is latent and can surface through development practice.}
The contract appears nowhere in the requirement, the function signature, or the dependency source. It is difficult for the agent to acquire this knowledge through searching or reading the repository code, as the preliminary experiments in Section~\ref{sec:motivation:exp} confirm. More broadly, tacit knowledge encompasses undocumented service contracts, workflow conventions, and business logic constraints. It is impractical to require that developers write out every such constraint. They are typically internalized through the accumulated experience of developers who have worked within the project over time~\cite{apimisuse, specmining}.

\textbf{Property 2: Tacit knowledge is structurally dispersed and propagates along dependency relations.} 
The contract does not reside in the target function alone. The factory and the downstream consumers that impose this obligation are structurally connected through the shared \texttt{llm\_router} node in the dependency graph. This dependency relation exists in both directions, as the factory writes to the router and the downstream stages read from it. Tacit knowledge manifests as a cross-component agreement spanning interface contracts between modules, implicit conventions in pipeline coordination, and consistency requirements across independently developed components. Such agreements naturally correspond to structural relations in the dependency graph.

\textbf{Property 3: The agent does not recognize its own tacit knowledge deficit.} The agent completed its trajectory with full confidence and submitted a structurally well-formed implementation that nonetheless failed every test. Analysis of the trajectory reveals that the agent at no point sought information about the two missing contracts, and its reasoning trace contains no indication that it perceived any gap in its understanding. When the missing knowledge is tacit, the agent's observable environment provides no cue that additional investigation is needed, leaving the agent with no basis on which to formulate the relevant queries.

These three properties motivate the core design of \ourapproach: to \textbf{extract} tacit knowledge through development practice rather than static reading (Property 1), to \textbf{organize} it on the code dependency graph so that it travels with the cross-component obligations it encodes (Property 2), and to \textbf{inject} it proactively whenever the agent touches the associated code, rather than relying on the agent to recognize its own knowledge deficit (Property 3).
\begin{figure*}[t]
  \centering
  \includegraphics[width=\textwidth]{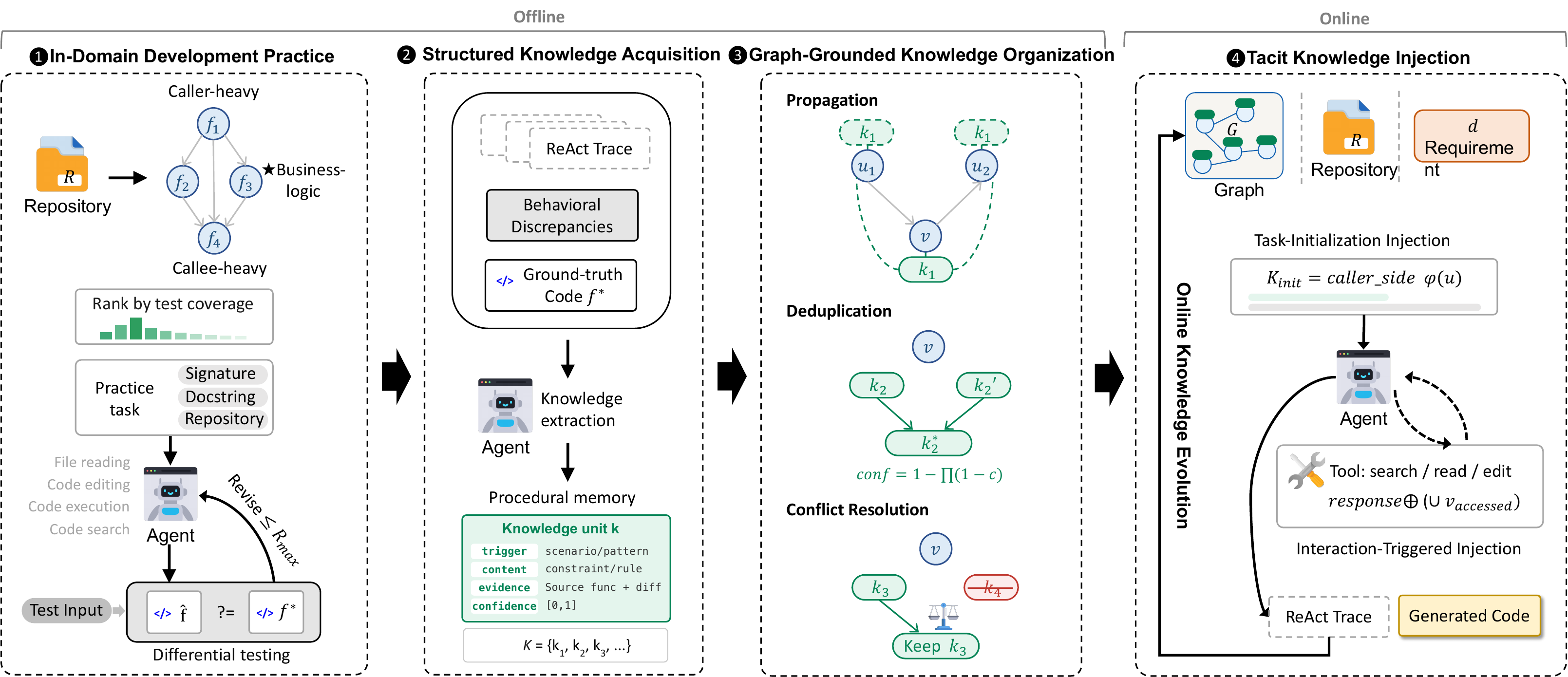}
  \caption{Overview of \ourapproach.}
  \label{fig:method}
\end{figure*}

\section{\ourapproach}
\label{sec:method}

\subsection{Problem Formulation}
\label{sec:formulation}

We consider the task of domain code generation within a specialized repository $\mathcal{R}$. Unlike general-purpose code generation, $\mathcal{R}$ is focused on a specific domain and contains project-specific conventions, constraints, and implementation patterns that are unlikely to appear in the model's pre-training corpus. The agent has full access to every file in $\mathcal{R}$, including source code, documentation, and configuration files.

A domain code generation task is defined as $\tau = (d, \mathcal{R})$, where $d$ is a natural language description of the required implementation and $\mathcal{R}$ is the target repository. Given $\tau$, an agent $\mathcal{A}$ works through the task in a manner analogous to a developer engaging with an unfamiliar codebase: it reasons about the requirements, explores the repository to understand relevant modules, reads related interfaces and conventions, writes code, and revises upon encountering issues. This process is formalized as a ReAct-style trajectory~\cite{react}:
\begin{equation}
\mathcal{T} = \langle \tau, h_1, a_1, o_1, h_2, a_2, o_2, \dots, h_T, a_T, o_T \rangle
\end{equation}
where at each step $i$, the agent produces a thought $h_i$ and an action $a_i$ (such as code search, file reading, or code editing), after which the environment returns an observation $o_i$. The final action $a_T$ submits the implementation $\hat{f}$. The task is considered resolved if and only if $\hat{f}$ passes all unit tests $\mathcal{U} = \{c_1, c_2, \dots, c_m\}$:
\begin{equation}
\text{Pass}(\hat{f}, \mathcal{U}) = \text{True}, \quad \text{i.e.,} \quad \forall j = 1..m, \ \text{Pass}(\hat{f}, c_j) = \text{True}
\end{equation}

\subsection{Overview}
\label{sec:overview}

PRAXIS overcomes the challenges posed by tacit knowledge through four progressive stages, as illustrated in Figure~\ref{fig:method}. 
The first two stages jointly address the knowledge extraction problem. \ding{182} In-Domain Development Practice (Section~\ref{sec:practice}) lets the agent perform real development work within $\mathcal{R}$ to surface tacit knowledge through behavioral discrepancies. \ding{183} Structured Knowledge Acquisition (Section~\ref{sec:acquisition}) then distills these discrepancies into structured, verifiable knowledge units. \ding{184} Graph-Grounded Knowledge Organization (Section~\ref{sec:organization}) addresses the representation problem by anchoring knowledge to code entities on a dependency graph and propagating it along dependency relations. \ding{185} Tacit Knowledge Injection (Section~\ref{sec:injection}) addresses the delivery problem by automatically surfacing relevant knowledge during downstream tasks without requiring the agent to recognize its own deficit. The first three stages are performed offline, while the fourth operates online during inference\footnote{All three offline stages strictly exclude evaluation test-set code, ensuring that no data leakage occurs~\cite{contamination, timetravel}. The practice functions are disjoint from the evaluation target functions, whose reference implementations are masked before the first stage begins, and the evaluation tests are never accessible.}.

\subsection{In-Domain Development Practice}
\label{sec:practice}

Let $\mathcal{F} = \{f_1, f_2, \dots, f_n\}$ denote the set of functions in $\mathcal{R}$, and let $\mathcal{G} = (V, E)$ be the dependency graph~\cite{pdg, callgraph} of $\mathcal{R}$, constructed via static analysis~\cite{Joern}. $V$ is the set of all functions in $\mathcal{R}$. $E \subseteq V \times V$ is a set of directed edges constructed in two steps: 1)~call dependencies: if function $u$ calls function $v$, an edge $(u, v)$ is added. 2)~data dependencies: if $u$ and $v$ share no call relation but $u$ consumes data produced by $v$, an edge $(u, v)$ is added. 
PRAXIS begins by selecting a practice set $\mathcal{F}_{\text{core}} \subseteq \mathcal{F}$ of functions on which the agent will perform development practice.

We identify candidate functions through three complementary strategies: (1)~business-logic functions that carry project-specific domain rules, identified by LLM-based semantic analysis that filters out generic utility code, as these functions are most likely to embed domain-specific tacit knowledge; (2)~callee-heavy functions with high in-degree in $\mathcal{G}$, representing core interfaces called by many modules, as tacit knowledge about these widely-used interfaces has the broadest downstream impact~\cite{chianti}; and (3)~caller-heavy functions with high out-degree in $\mathcal{G}$, which invoke many other modules and whose correct implementation requires understanding multiple interface contracts, as these functions typically involve multi-hop tacit knowledge and business logic pipelines. The union of these three strategies forms the candidate pool $\mathcal{F}_{\text{cand}} \subseteq \mathcal{F}$.

For each $f_i \in \mathcal{F}_{\text{cand}}$ with ground-truth implementation $f_i^*$, we construct a practice task as follows. LLM generates a natural language requirement description $d_i$ and a diverse set of test inputs $\mathcal{X}_i = \{x_1, \dots, x_l\}$ from $f_i^*$~\cite{toga, codamosa, chattester, intut, aster}.
To ensure the quality of each practice task, we measure the line coverage of $\mathcal{X}_i$ over $f_i^*$ and retain only candidates exceeding 80\%, guaranteeing that the generated test inputs comprehensively exercise the function's behavior and that differential testing can reliably detect discrepancies. The resulting functions form the practice set $\mathcal{F}_{\text{core}}$. For each $f_i \in \mathcal{F}_{\text{core}}$, the function body is removed, retaining only the signature $s_i$ and docstring. The agent generates an implementation:
\begin{equation}
\hat{f}_i \sim \mathcal{A}(\cdot \mid s_i, d_i, \mathcal{R})
\end{equation}
Each practice session is conducted independently.

The generated implementation is evaluated through differential testing~\cite{mckeeman1998, csmith, oraclesurvey, opdiffer, dllens}. Each test input $x_j \in \mathcal{X}_i$ is passed to both $\hat{f}_i$ and $f_i^*$, producing a set of behavioral discrepancies:
\begin{equation}
\mathcal{D}_i = \{(x_j, \hat{f}_i(x_j), f_i^*(x_j)) \mid \hat{f}_i(x_j) \neq f_i^*(x_j), \ x_j \in \mathcal{X}_i\}
\end{equation}
Each discrepancy in $\mathcal{D}_i$ is returned to the agent as an observation within its practice trajectory, providing concrete feedback for the subsequent refinement round. The agent revises its implementation as $\hat{f}_i^{(r+1)} \leftarrow \mathcal{A}(\hat{f}_i^{(r)}, \mathcal{D}_i^{(r)})$, where $r$ is the refinement round. This cycle repeats until $\mathcal{D}_i = \emptyset$ or a maximum of $R_{\max}$ rounds is reached. 
We denote the complete practice trajectory for function $f_i$, encompassing the agent's code exploration, implementation attempts, error diagnosis, and iterative revisions, as $\mathcal{T}_i^{p}$.

\subsection{Structured Knowledge Acquisition}
\label{sec:acquisition}

For each practiced function $f_i$, we construct a structured diff: 
\begin{equation} 
    \Delta_i = \langle \mathcal{T}_i^{p}, \mathcal{D}_i, f_i^* \rangle 
\end{equation} 
juxtaposing the agent's practice trajectory (which includes all implementation attempts) and the behavioral discrepancies. If the agent fails to resolve all discrepancies after multiple attempts, we provide the ground-truth implementation $f_i^*$ for comparison, enabling knowledge extraction from the differences.

Given $\Delta_i$, the agent extracts a set of knowledge units $\mathcal{K}_i$ grounded in these concrete differences rather than through free-form summarization. Each unit $k \in \mathcal{K}_i$ is stored as a piece of procedural memory~\cite{generativeagents, expel, awm, amem, procmem}, represented as a four-field tuple:
\begin{equation}
k = (\textit{trigger}, \textit{content}, \textit{evidence}, \textit{confidence})
\end{equation}
where \textit{trigger} describes the scenario under which this knowledge should be activated, supporting the downstream injection stage in determining when a unit is relevant; \textit{content} states the specific constraint or convention; \textit{evidence} records the source function and diff segment from which it was derived, ensuring traceability and verifiability; and \textit{confidence} $\in [0, 1]$ indicates the reliability of this knowledge unit, scored by the LLM based on two factors, i.e., whether the agent's implementation passed differential tests during practice, and the specificity and clarity of the supporting evidence. The full set of knowledge units across all practiced functions is $\mathcal{K} = \bigcup_{f_i \in \mathcal{F}_{\text{core}}} \mathcal{K}_i$.

\subsection{Graph-Grounded Knowledge Organization}
\label{sec:organization}

Each knowledge unit $k \in \mathcal{K}$ is anchored to its associated code entity through a mapping $\phi: V \rightarrow 2^{\mathcal{K}}$. For a unit extracted from practicing function $f_i$, we set $\phi(v_i) \leftarrow \phi(v_i) \cup \{k\}$, where $v_i \in V$ is the corresponding node in $\mathcal{G}$.

\textbf{Knowledge Propagation.}
Certain knowledge units have implications beyond their source entity. We propagate knowledge along edges in both directions, as a function's constraints can affect every function that depends on it, and conversely, the way a function is used can impose requirements on the functions it depends on. For an edge $(u, v) \in E$, each $k \in \phi(v)$ deemed relevant to $u$ by an LLM-based relevance assessment yields a propagated version $k'$, with an updated trigger reflecting $u$'s perspective, added to $\phi(u)$; symmetrically, each $k \in \phi(u)$ deemed relevant to $v$ yields $k'$ reflecting $v$'s perspective, added to $\phi(v)$. Propagation proceeds iteratively until convergence or a maximum depth is reached.

\textbf{Deduplication and Conflict Resolution.}
For each node $v$, semantically equivalent units within $\phi(v)$ are merged. Since independently extracted units that express the same knowledge provide mutual verification, their aggregated confidence should be increased. We compute the merged confidence as:
\begin{equation}
\textit{confidence}(k_{\text{merged}}) = 1 - \prod_{k \in S}(1 - \textit{confidence}(k))
\end{equation}
where $S$ is the set of equivalent units. When contradictory units are detected, an LLM adjudicator~\cite{llmjudge} considers both units' evidence fields and confidence scores to determine which to retain. The result is a knowledge-annotated dependency graph $\mathcal{G_K} = (V, E, \phi)$.

\subsection{Tacit Knowledge Injection}
\label{sec:injection}
Since the agent is unaware of which tacit knowledge it lacks, we proactively inject it during the agent's development workflow.
The injection process is divided into two stages, Task-Initialization Injection and Interaction-Triggered Injection, both of which filter out low-quality units through a filtered mapping $\phi_\theta(v) = \{k \in \phi(v) \mid \textit{confidence}(k) \geq \theta\}$. 

\textbf{Task-Initialization Injection.}
When a new task $\tau$ is received, the system identifies the target code entity $v_\tau \in V$ from the task description and retrieves callers of $v_\tau$: $V_{\text{caller}} = \{u \mid (u, v_\tau) \in E\}$. The associated knowledge is collected and injected into the agent's initial context:
\begin{equation}
\mathcal{K}_{\text{init}} = \bigcup_{u \in V_{\text{caller}}} \phi_\theta(u)
\end{equation}
This provides the agent with caller-side constraints, including input requirements, return value conventions, and usage expectations, before it begins writing code.

\textbf{Interaction-Triggered Injection.}
During the agent's coding process, the injection mechanism is embedded into the agent's existing tools. When the agent performs code search, file reading, or code editing, the system identifies the accessed code entities $V_{\text{accessed}}$. The associated knowledge is appended to the tool's return results:
\begin{equation}
\textit{response}' = \textit{response} \oplus \bigg( \bigcup_{v \in V_{\text{accessed}}} \phi_\theta(v)\bigg)
\end{equation}
where $\oplus$ denotes appending knowledge units to the original response.

\textbf{Online Knowledge Evolution.}
When the agent encounters new tasks on the same codebase, tacit knowledge can be extracted from the online task trajectories through the same structured knowledge acquisition pipeline and organized onto $\mathcal{G_K}$ via anchoring and propagation. The confidence of existing knowledge units is updated based on task outcomes. For a unit $k$ that is hit and contributed to a successful task, its confidence is reinforced as $\textit{confidence}(k) \leftarrow 1 - \beta(1 - \textit{confidence}(k))$; for a unit that is hit but the task failed, its confidence is decayed as $\textit{confidence}(k) \leftarrow \beta \cdot \textit{confidence}(k)$, where $\beta \in (0, 1)$.
\section{Experiments}
To evaluate the effectiveness of \ourapproach, we conduct a comprehensive large-scale study. We aim to answer the following research
questions (RQs):

\smallskip
\noindent\textit{Effectiveness and component contribution.}
\begin{rqlist}
  \item \textbf{RQ1 (Overall Effectiveness).} How does \ourapproach compare against state-of-the-art agents and existing experience- and skill-based methods on domain code generation?
  \item \textbf{RQ2 (Ablation Study).} How does each of components of \ourapproach contribute to its effectiveness?
\end{rqlist}

\smallskip
\noindent\textit{Generalization and adaptability.}
\begin{rqlist}
\item \textbf{RQ3 (Across Agent Frameworks and Base Models).}
Does \ourapproach generalize across different agent frameworks and
LLMs of different scales and families?
  \item \textbf{RQ4 (Across Benchmarks and Tasks).} Does \ourapproach transfer to a new benchmark and task type, i.e., repository-level bug fixing?
\end{rqlist}

\smallskip
\noindent\textit{Practice scaling and continual evolution.}
\begin{rqlist}
  \item \textbf{RQ5 (Practice Scaling).} How does performance change as the amount of offline development practice increases?
  \item \textbf{RQ6 (Online Knowledge Evolution).} Can \ourapproach continually accumulate knowledge online during real task execution?
\end{rqlist}

\smallskip
\noindent\textit{Further analysis.}
\begin{rqlist}
  \item \textbf{RQ7 (Cost Analysis).} What is the token cost of \ourapproach relative to baselines?
  \item \textbf{RQ8 (Knowledge Quality and Typology).} What kind of tacit knowledge does \ourapproach extract, and is it of high quality?
\end{rqlist}

\subsection{Experimental Setup}

\paragraph{Benchmarks}
We select benchmarks that are designed to evaluate domain specialization in real-world software development rather than general-purpose code generation.
\textbf{KoCo-Bench}~\cite{kocobench} is our primary benchmark. It spans multiple emerging
domains (including Reinforcement Learning, Agent, RAG, Model Optimization) across 11 software frameworks and 25 real-world
projects, and provides code generation tasks, each backed by rigorous test suites.
To assess whether \ourapproach transfers to other benchmarks and tasks, we further adopt \textbf{AInsteinBench}~\cite{ainsteinbench}, a large-scale domain
benchmark for evaluating LLM agents in scientific computing development.
AInsteinBench contains bug-fixing tasks derived from maintainer-authored
pull requests across six production scientific codebases grouped into five scientific domains:
Numerical Relativity (Einstein Toolkit, AMReX), Quantum Information (Qiskit),
Molecular Dynamics (OpenMM), Quantum Chemistry (PySCF), and Cheminformatics (RDKit).
All tasks are grounded in executable environments with tests.

\paragraph{Baselines}
We focus on agent-based approaches, which serve as unified systems
that integrate retrieval, agentic search, code execution, static
analysis, and other techniques, and have been shown to achieve the
best performance on domain code generation~\cite{kocobench,ainsteinbench}.
Within this scope, we compare \ourapproach against three categories of methods.
1)~\emph{Representative LLM Agents}, including
OpenHands~\cite{openhands}, SWE-Agent~\cite{sweagent},
OpenCode~\cite{opencode}, and OpenCollab~\cite{hsu2026opencollab,DBLP:journals/tosem/DongJJL24}, each of which employs a distinctly designed
set of tools for repository exploration and agentic search.
2)~\emph{Experience-based evolution methods} accumulate problem-solving
experience from agent trajectories and retrieve relevant experience to guide
new tasks. We compare against the recent SOTA method in this
category, SWE-Exp~\cite{sweexp}.
3)~\emph{Skill-based evolution methods} distill reusable skills (i.e., transferable action patterns and code strategies) from agent
trajectories and invoke them at inference time. We compare against the recent
SOTA method in this category, Trace2Skill~\cite{trace2skill}.
To ensure a fair comparison, all methods use the same base model and
receive identical inputs at the start of each task, consisting of the task
requirement and the initial repository state. The repository is equally accessible to
all methods, and how it is utilized is determined by each individual method.
The experience-based and skill-based methods further share the same agent
framework as \ourapproach.

\begin{table*}[t!]
\caption{Overall effectiveness on KoCo-Bench. \textbf{Bold} indicates the best result per column; \underline{underline} indicates the second best. Average is weighted by the number of tasks per domain.}
\label{tab:rq1}
\centering
\footnotesize
\setlength{\tabcolsep}{3.8pt}
\begin{tabular}{@{}l rr rr rr rr rr@{}}
\toprule
& \multicolumn{2}{c}{\makecell{RL \\Domain}} & \multicolumn{2}{c}{\makecell{RAG \\Domain}} & \multicolumn{2}{c}{\makecell{Agent \\Domain}} & \multicolumn{2}{c}{\makecell{Model Optimization \\Domain}} & \multicolumn{2}{c}{Average} \\
\cmidrule(lr){2-3} \cmidrule(lr){4-5} \cmidrule(lr){6-7} \cmidrule(lr){8-9} \cmidrule(lr){10-11}
Method & Pass@1 & AvgPassRatio & Pass@1 & AvgPassRatio & Pass@1 & AvgPassRatio & Pass@1 & AvgPassRatio & Pass@1 & AvgPassRatio \\
\midrule
Base Model & 7.55 & 23.97 & 0.00 & 0.00 & 13.46 & 37.00 & 16.67 & 38.30 & 10.69 & 29.65 \\
\midrule
OpenHands & 18.87 & 49.86 & 12.50 & 22.92 & 19.23 & 42.25 & 22.22 & 48.13 & 19.08 & 44.96 \\
SWE-Agent & 18.87 & 50.96 & \textbf{50.00} & \underline{50.00} & 28.85 & 43.73 & \underline{27.78} & 49.84 & 25.96 & 47.88 \\
OpenCode & \underline{20.76} & \textbf{56.61} & \underline{37.50} & 43.75 & 26.92 & 49.34 & 22.22 & 52.60 & 24.43 & \underline{52.39} \\
OpenCollab & 20.75 & 48.48 & \underline{37.50} & 45.83 & \underline{34.62} & \underline{51.26} & 22.22 & \underline{54.79} & \underline{27.48} & 50.29 \\
\midrule
SWE-Exp & 6.15 & 13.68 & 0.00 & 0.00 & 7.69 & 24.86 & 16.67 & 39.56 & 7.83 & 20.84 \\
\midrule
Trace2Skill & 16.98 & 54.21 & \underline{37.50} & 40.62 & 30.76 & 50.69 & 27.77 & 50.11 & 25.19 & 51.42 \\
\midrule
\ourapproach & \textbf{22.64} & \underline{56.46} & \textbf{50.00} & \textbf{55.36} & \textbf{36.54} & \textbf{54.67} & \textbf{38.89} & \textbf{58.83} & \textbf{32.06} & \textbf{56.01} \\
\bottomrule
\end{tabular}
\end{table*}

\paragraph{Implementation Details}
We use DeepSeek-V3.2~\cite{deepseekv3_2} as the base model and OpenHands~\cite{openhands} as the
agent framework for all experiments unless otherwise specified. To evaluate
generalization across agent frameworks, we additionally run experiments with
SWE-Agent~\cite{sweagent}. We report two metrics: Pass@1~\cite{codex}, the proportion
of tasks whose generated implementation passes all unit tests, and AvgPassRatio, the average proportion of test cases passed per
task, which provides a fine-grained measure of partial correctness. Hyperparameters are set as follows: $R_{\max} = \text{3}$ refinement
rounds for differential testing, a maximum propagation depth of 4 hops on the
dependency graph, a confidence threshold $\theta = \text{0.7}$ for knowledge
injection, and an update factor $\beta = \text{0.8}$ for online evolution. All experiments use greedy decoding with a temperature of 0 to ensure deterministic and reproducible results.

\subsection{Overall Effectiveness and Component Contribution}

\paragraph{RQ1: Overall Effectiveness}
Table~\ref{tab:rq1} presents the overall results\footnote{To confirm that PRAXIS yields reliable improvements over all baselines, we conducted a paired $t$-test, and the results show that the differences on AvgPassRatio are statistically significant ($p < 0.05$).}. \ourapproach achieves
the highest average Pass@1 and AvgPassRatio across all domains,
outperforming the second-best method by a relative improvement of
16.7\% on Pass@1. Compared with existing representative open-source
agent frameworks, \ourapproach consistently surpasses OpenHands,
SWE-Agent, OpenCode, and OpenCollab despite these agents being equipped with their
own distinctly designed repository exploration and search capabilities.
\ourapproach also outperforms both the experience-based and skill-based
evolution methods, demonstrating that organizing knowledge on the code
dependency graph and proactively delivering it at the point of code
interaction provides a significant competitive advantage over flat
retrieval from experience or skill banks. Trace2Skill, the strongest
evolution-based baseline, delivers more consistent results than
SWE-Exp, yet it still falls short of \ourapproach across domains. In
contrast, the experience-based method SWE-Exp exhibits notably weak
performance, falling below even the raw base model on average Pass@1.
We attribute this to two factors: SWE-Exp lacks a mechanism to assess
the quality of the knowledge it extracts, meaning it utilizes
everything indiscriminately, and the experiences it accumulates tend to
be overly general, lacking the specificity required to capture the
domain conventions and implicit contracts that govern correct
implementation in these codebases.

\paragraph{RQ2: Ablation Study}

\begin{table}[t]
  \caption{Results of Ablation Study.}
  \label{tab:ablation}
  \centering
  \footnotesize
  \begin{tabular}{lcc}
    \toprule
    Method & Avg. Pass@1 & Avg. AvgPassRatio \\
    \midrule
    Base Model              & 10.69          & 29.65          \\
    OpenHands               & 19.08          & 44.96          \\
    \ourapproach             & \textbf{32.06} & \textbf{56.01} \\
    \midrule
    w/o Development Practice & 27.48 & 51.17   \\
    w/o Procedural Memory    & 29.01 & 50.69      \\
    w/o Graph Organization   & 28.25          & 48.80          \\
    w/o Proactive Injection  & 28.24          & 53.36 \\
    \bottomrule
  \end{tabular}
\end{table}

To isolate the contribution of the four stages, we construct four variants
by replacing or removing one component at a time.
1)~\textbf{w/o Development Practice} replaces practice-based knowledge
extraction with direct extraction from source code.
2)~\textbf{w/o Procedural Memory} replaces the structured four-tuple
representation with flat text descriptions.
3)~\textbf{w/o Graph Organization} disables graph-based propagation,
deduplication, and conflict resolution.
4)~\textbf{w/o Proactive Injection} replaces interaction-triggered
injection with embedding-based retrieval.
Table~\ref{tab:ablation} reports the results. Removing development
practice causes the largest Pass@1 drop (32.06 $\to$ 27.48),
confirming that tacit knowledge cannot be recovered through direct
source code analysis and that simulating human development workflows is
essential for surfacing it. The remaining three components each
contribute meaningfully: graph-based organization enables knowledge to
reach the code entities it can influence while eliminating redundancies
and conflicts, proactive injection surfaces knowledge to the agent
without requiring it to determine what to retrieve on its own, and the
structured procedural memory format provides a more efficient
representation for expressing and activating knowledge. All four
stages work together to produce the full performance gain.

\begin{table}[t]
  \caption{Generalization across agent frameworks and base models on
  KoCo-Bench.}
  \label{tab:generalization}
  \centering
  \resizebox{0.49\textwidth}{!}{
  \begin{tabular}{ll cc cc}
    \toprule
    & & \multicolumn{2}{c}{Avg. Pass@1}
      & \multicolumn{2}{c}{Avg. AvgPassRatio} \\
    \cmidrule(lr){3-4} \cmidrule(lr){5-6}
    Agent & Base Model
      & Baseline & \ourapproach
      & Baseline & \ourapproach \\
    \midrule
    OpenHands & GPT-5.5-20260424
      & 16.03 & \textbf{42.75}
      & 27.72 & \textbf{60.44} \\    
    OpenHands & Qwen3.6-Plus
      & 25.95 & \textbf{27.48}
      & 45.71 & \textbf{51.31} \\
    OpenHands & DeepSeek-V3.2
      & 19.08 & \textbf{32.06}
      & 44.96 & \textbf{56.01} \\      
    SWE-Agent & DeepSeek-V3.2
      & 25.96 & \textbf{35.12}
      & 47.88 & \textbf{56.85} \\
    \bottomrule
  \end{tabular}
  }
\end{table}

\begin{table*}[t!]
  \centering
  \caption{Results on AInsteinBench. \%Resolved reports the proportion
  of bug-fixing tasks fully resolved. Difficulty stratifies results by
  task complexity as annotated in the benchmark.}
  \label{tab:ainsteinbench}
  \footnotesize
  \setlength{\tabcolsep}{4pt}
  \begin{tabular}{lc ccc ccccc}
    \toprule
    \multirow{2}{*}{Method}
      & \multirow{2}{*}{\%Resolved$\uparrow$}
      & \multicolumn{3}{c}{Difficulty}
      & \multicolumn{5}{c}{Scientific Domain} \\
    \cmidrule(lr){3-5} \cmidrule(lr){6-10}
      & & Hard & Medium & Easy
      & \makecell{Numerical\\Relativity}
      & \makecell{Quantum\\Information}
      & \makecell{Molecular\\Dynamics}
      & \makecell{Quantum\\Chemistry}
      & \makecell{Chem-\\informatics} \\
    \midrule
    OpenHands
      & 27.2 & 19.1 & 29.2 & 33.8
      & 23.7 & 51.9 & 33.3 & 31.9 & 0.0 \\
    \ourapproach
      & \textbf{31.2} & \textbf{26.6} & \textbf{33.3} & \textbf{35.1}
      & \textbf{28.8} & \textbf{55.6} & \textbf{35.2} & \textbf{36.2} & \textbf{5.4} \\
    \bottomrule
  \end{tabular}
\end{table*}

\subsection{Generalization and Adaptability}

\paragraph{RQ3: Across Agent Frameworks and Base Models}
Table~\ref{tab:generalization} reports the results of pairing
\ourapproach with two agent frameworks and three base models from
different families. Across all four configurations, \ourapproach
consistently improves over the baseline on both metrics, confirming
that its benefits generalize across agent frameworks and LLMs of
varying scales and pretraining backgrounds. Comparing the two
frameworks on the same base model (DeepSeek-V3.2), \ourapproach yields
substantial gains on both OpenHands and SWE-Agent, indicating that the
knowledge extraction and injection pipeline is not tied to any
particular agent architecture. Among base models, the improvement is
most pronounced on GPT-5.5. Models with stronger general capabilities
but less domain-specific coverage in their pretraining corpus benefit
most once the missing tacit knowledge is supplied.

\paragraph{RQ4: Across Benchmarks and Tasks}
To assess whether \ourapproach transfers beyond domain code generation,
we evaluate on AInsteinBench, a domain benchmark of repository-level bug
fixing across six production scientific computing codebases. Bug fixing
differs from code generation in that the agent must first
localize the fault and then produce a correct patch, making tacit
knowledge about project-specific conventions equally critical. We apply
\ourapproach to AInsteinBench without any modification to the pipeline.
Table~\ref{tab:ainsteinbench} reports the results. \ourapproach
improves the overall resolution rate and
achieves gains across all three difficulty levels.
Performance also improves consistently across all scientific domains,
notably in numerical relativity (23.7\% $\to$ 28.8\%) and quantum
chemistry (31.9\% $\to$ 36.2\%), while also enabling the agent to
resolve cheminformatics tasks that the baseline fails entirely on
(0.0\% $\to$ 5.4\%). The consistent gains across a new benchmark, a
different task type, and diverse scientific domains confirm the
generalizability of \ourapproach.

\subsection{Practice Scaling and Continual Evolution}

\begin{figure}[t]
  \centering
  \includegraphics[width=0.95\linewidth]{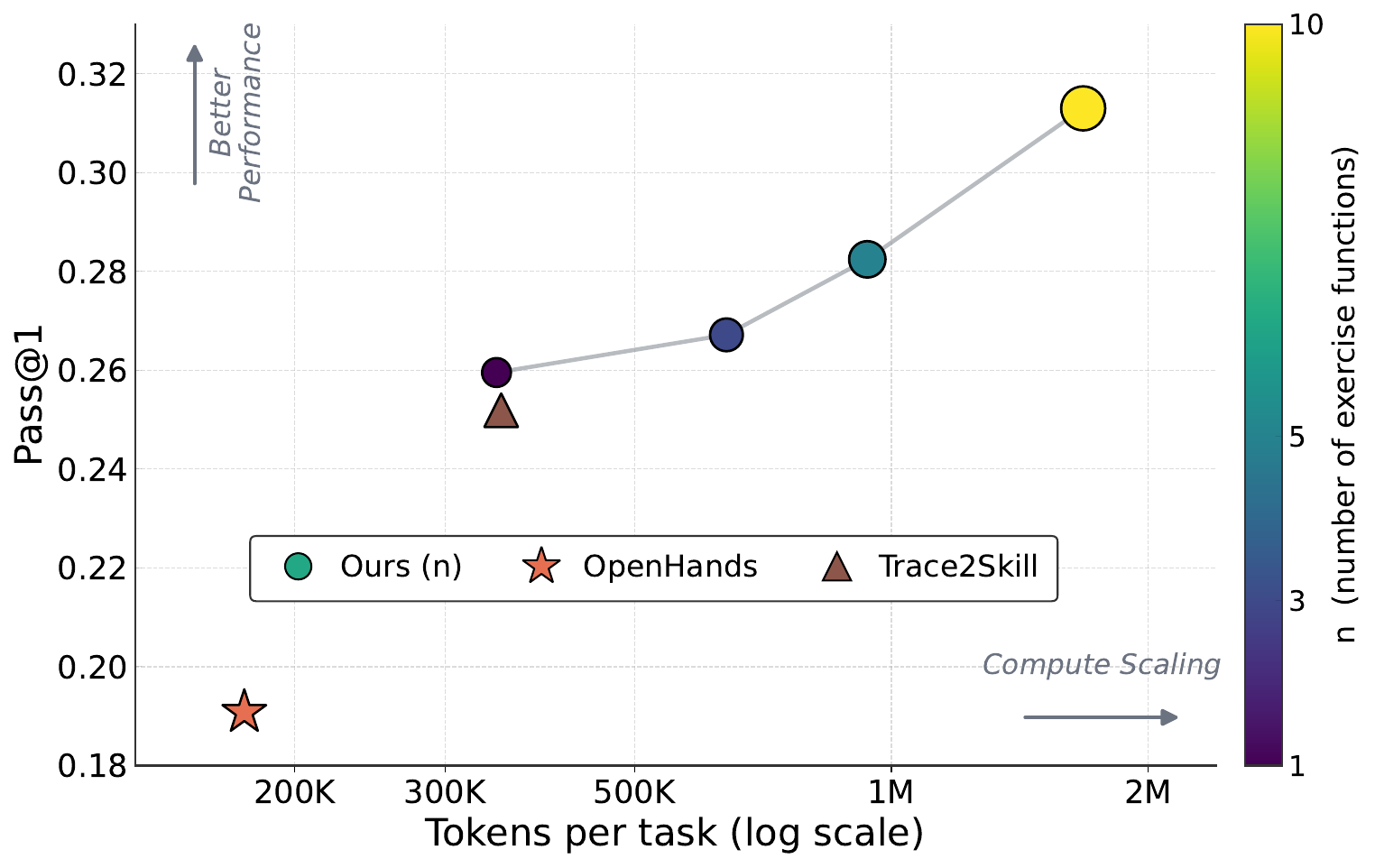}
  \caption{Results of performance scaling and token cost.}
  \label{fig:practice_scaling_cost}
\end{figure}

\paragraph{RQ5: Practice Scaling}
To explore whether the effort invested in offline practice translates
proportionally into performance gains, we vary the number of practiced
functions from 1 to 10 and measure the resulting Pass@1.
Figure~\ref{fig:practice_scaling_cost} shows the results. \ourapproach
yields a steady performance gain as the practice set grows. Even a
single practiced function already lifts Pass@1 well above the
OpenHands baseline and the strongest evolution-based baseline Trace2Skill, and each subsequent increase in practice volume
brings further improvement. This confirms that the
tacit knowledge extracted through development practice is cumulative, contributing to downstream task performance.

\paragraph{RQ6: Online Knowledge Evolution}

\begin{figure}[t]
  \centering
  \includegraphics[width=0.95\linewidth]{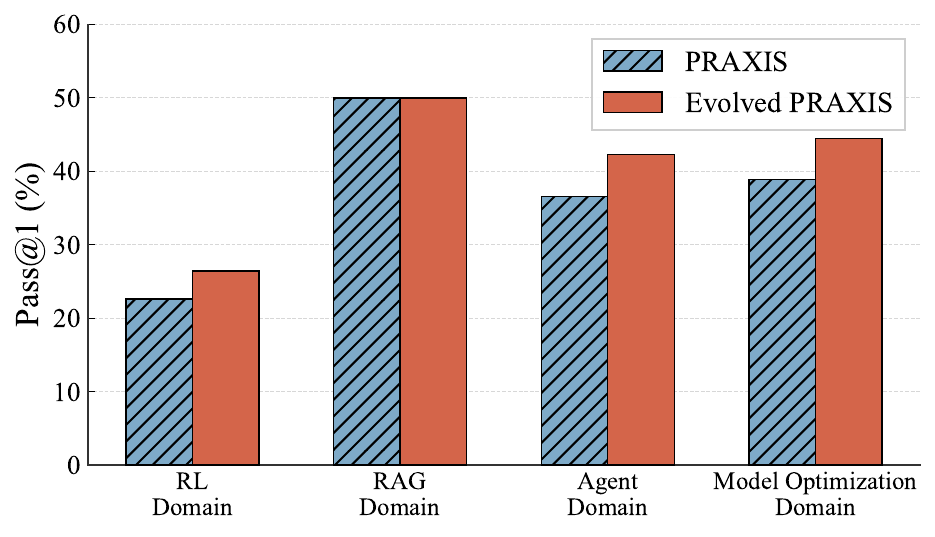}
  \caption{Results of Online Knowledge Evolution with \ourapproach.}
  \label{fig:knowledge_evolution}
\end{figure}

To evaluate whether \ourapproach can continually accumulate knowledge
during real task execution, we process evaluation tasks sequentially
within each domain. After each task, knowledge is extracted from the
agent's trajectory through the same structured acquisition pipeline
and organized onto the dependency graph for use in subsequent tasks. Evolution is conducted within each domain independently.
Figure~\ref{fig:knowledge_evolution} reports the results. Online
evolution yields improvements across most domains, while the RAG
domain remains stable. This demonstrates that \ourapproach can
continuously accumulate knowledge and improve performance during
deployment without regression.

\subsection{Further Analysis}

\paragraph{RQ7: Cost Analysis}
Figure~\ref{fig:practice_scaling_cost} plots the token cost against Pass@1 for
\ourapproach and the baselines.
Compared with OpenHands, \ourapproach does introduce additional token
cost due to the offline practice phase, but this overhead translates
into a substantial performance gain. Compared with Trace2Skill, the
strongest evolution-based baseline, \ourapproach at $n=1$ already achieves higher
Pass@1 under a comparable token budget, confirming that our approach is cost-effective even at its lightest
configuration. Beyond this, \ourapproach exhibits a scaling property analogous to
test-time scaling, where Pass@1 improves steadily as the number of
practiced functions increases. This means that stronger downstream
performance can be achieved simply by investing additional
computational resources in practice, a property widely recognized as
highly desirable and one that baseline methods such as Trace2Skill do
not possess. Moreover, the cost of \ourapproach is concentrated
entirely in the offline practice phase, which only needs to be
performed once per codebase and is then amortized across all
subsequent tasks within that repository.

\paragraph{RQ8: Knowledge Quality and Typological Analysis}
We examine the extracted tacit knowledge to understand what kind it takes, and whether it is reliable. We recruit two volunteer engineers, each with over two years of development experience, to conduct the following analysis.

We randomly sample 75 knowledge units from the knowledge base and the evaluators categorize each unit. 
The analysis reveals five distinct types, i.e., \emph{business rules}, \emph{API patterns}, \emph{interface contracts}, \emph{error handling conventions}, and \emph{others}.
\emph{Business rules}, encompassing reward function conventions, algorithmic
constraints, and verification protocols unique to the target project,
are the most prevalent (34.7\%). \emph{API patterns} account for 24.0\%, followed by \emph{interface contracts} encoding undocumented assumptions about input types, shapes,
and return conventions (22.7\%), and \emph{error handling conventions}
covering project-specific decisions about exception propagation and
default returns (14.7\%). The remaining 4.0\% reflect non-behavioral
stylistic preferences. All major categories encode knowledge that is difficult to
extract directly from the source code, confirming their nature as
tacit knowledge.

The evaluators also rate each sampled unit on three dimensions using a
5-point scale: \emph{content truthfulness}, \emph{trigger accuracy},
and \emph{evidence traceability}. As shown in Table~\ref{tab:quality},
all three dimensions receive high scores (mean 4.28, 4.24, and 3.62),
confirming the overall quality of the extracted knowledge. To
assess whether the automatically assigned confidence scores align with
human judgments, we compute the Spearman rank correlation between each
unit's confidence and its average human quality rating, obtaining
$\rho = 0.75$ ($p < 0.001$), confirming that the confidence score
provides a meaningful signal for quality-based filtering.
We further manually inspect the trajectories of 50 randomly sampled tasks and
find that 82\% of them utilize the extracted knowledge, of which
87.8\% produce a meaningful contribution to the task.

\begin{table}[t]
  \caption{Human quality assessment of extracted knowledge units
  (5-point scale) and confidence reliability.}
  \label{tab:quality}
  \centering
  \resizebox{0.44\textwidth}{!}{
  \begin{tabular}{lcc}
    \toprule
    Dimension & Mean & Median \\
    \midrule
    Content Truthfulness    & 4.28 & 5 \\
    Trigger Accuracy        & 4.24 & 4 \\
    Evidence Traceability   & 3.62 & 4 \\
    \midrule
    Confidence Reliability (Spearman $\rho$)
      & \multicolumn{2}{c}{$0.75$ ($p < 0.001$)} \\
    \bottomrule
  \end{tabular}
  }
\end{table}
\section{Related Work}
\label{sec:related}

\subsection{Domain Code Generation}

LLM-based agents have achieved strong performance on general software engineering benchmarks such as SWE-bench~\cite{swebench}, where leading systems resolve the majority of real-world GitHub issues. However, domain-specific software development remains a significant challenge. Unlike general programming~\cite{deepseekcoderv2, qwen3_coder, starcoder2, DBLP:journals/tosem/JiangDWFSLJJ24}, domain code generation requires specialized knowledge that is absent from the model's pre-training corpus, including proprietary APIs, domain-specific constraints, and project-level implementation conventions~\cite{kocobench}. Recent benchmarks such as KoCo-Bench~\cite{kocobench} reveal that even state-of-the-art LLMs achieve remarkably low accuracy on domain code generation tasks, a finding corroborated by other independent studies~\cite{domaincodebench}. Existing domain specialization methods have been proposed to address this gap. Parameter-updating methods such as SFT~\cite{dong2024sft} train models on domain data but are constrained by the high cost of curating labeled data, the risk of catastrophic forgetting, and a tendency toward shallow pattern learning. Inference-time approaches such as RAG~\cite{lewis2020rag}, kNN-LM~\cite{khandelwal2020knnlm}, and DomCoder~\cite{domcoder} inject domain knowledge during generation through retrieval or token-level fusion, but struggle with complex reasoning over fragmented and implicit domain knowledge. Collaborative approaches such as large-small model collaboration~\cite{yu2025collaboration} combine fine-tuned domain experts with general-purpose models, yet remain limited by the quality of domain-specific training data. These methods all yield only marginal and inconsistent improvements in practice~\cite{kocobench}. To date, agent-based systems such as Claude Code show the best performance on domain code generation, yet the problem remains far from solved.

Given that agent-based systems have demonstrated the strongest results on domain code generation, we build upon this foundation. PRAXIS complements existing methods by targeting a fundamentally different source of the domain gap. It draws on tacit knowledge that exists only in developers' minds and remains inaccessible to both learning-based and retrieval-based approaches.

\subsection{LLM Agents for Code Generation}

LLM-based agents have emerged as a powerful approach for automated software engineering~\cite{hsu2026opencollab,liu2026multiagent}. Self-Collaboration~\cite{DBLP:journals/tosem/DongJJL24} introduced a multi-agent framework where LLM instances assume distinct roles such as analyst, coder, and tester, demonstrating the potential of collaborative agent workflows for code generation. SWE-Agent~\cite{sweagent} subsequently designed agent-computer interfaces tailored for repository navigation, and AutoCodeRover~\cite{autocoderover} combined code search with program analysis. Agentless~\cite{agentless} proposed decomposing issue resolution into structured localization and repair without persistent agent state. HyperAgent~\cite{hyperagent} introduced a generalist multi-tool agent for repository-level tasks. OpenHands~\cite{openhands} provides an open platform for building generalist software development agents built on the ReAct architecture, interleaving reasoning with tool invocation in an iterative loop. In parallel, a growing body of work improves agent capability through training and alignment~\cite{liu2026sprout}. LingMa~\cite{lingma} introduces a development-process-centric language model for automated software improvement. SWE-Gym~\cite{swegym} provides executable training environments with human-curated test cases. SWE-RL~\cite{swerl} applies reinforcement learning with verifiable rewards to advance agent reasoning. SEAlign~\cite{sealign} aligns code models with real-world agentic workflows via preference optimization on critical action steps. SWE-Synth~\cite{swesynth} synthesizes verifiable bug-fix data with process-aware repair traces for supervised fine-tuning. SWE-Smith~\cite{swesmith} scales automated data collection for agent training. These methods have improved agents' general capabilities in instruction following, tool use, and reasoning.

All these efforts improve what the agent can do in general, whereas PRAXIS provides what the agent needs to know about a specific codebase. PRAXIS is training-free and can be stacked on top of any existing agent for additional domain-specific gains.

\subsection{Experience and Skill Evolution for Code Agents}

Recent work has explored how agents can improve through accumulated experience and reusable skills. On the experience side, AgentKB~\cite{agentkb} and AutoRefine~\cite{autorefine} construct an experience base from agent problem-solving trajectories and retrieve relevant entries for new problems, though they have not been applied to code generation tasks. Within software engineering, iterative experience refinement~\cite{iterative_exp} applies trajectory-based learning to software development agents, and LLMs as Continuous Learners~\cite{llm_continuous} accumulates experience from past software issues to improve defective code reproduction. SWE-Exp~\cite{sweexp} mines resolved software issues and retrieves relevant experiences at inference time to guide future issue resolution, representing the most direct application of experience-based learning to software engineering tasks. On the skill side, Kimi-Dev~\cite{kimidev} uses supervised fine-tuning to distill agentic workflows into a skill prior for SWE agents, requiring a full training pipeline. SICA~\cite{sica} presents a self-improving coding agent that iteratively refines its capabilities through repeated task execution. CodeSkill~\cite{codeskill} extracts multi-granularity procedural skills from coding-agent trajectories and maintains a skill bank, training the skill extraction policy via reinforcement learning with a hybrid reward combining rubric-based quality feedback and verifiable execution outcomes. Trace2Skill~\cite{trace2skill} distills trajectory-local lessons into transferable agent skills through prompt-based extraction and retrieval, offering a training-free alternative applicable across different agent settings.

Our method differs from these experience- and skill-based approaches along three dimensions. First, after acquiring experiences or skills, existing methods perform a global injection by retrieving relevant entries based on task-level similarity and supplying them as a monolithic block in the agent's context, whereas PRAXIS performs a distributed injection, binding each unit to a specific code entity and delivering it precisely at the point of interaction. Second, existing methods organize experiences and skills as flat collections indexed by text embeddings, whereas PRAXIS organizes them on the code dependency graph, capturing multi-hop constraints that flat retrieval cannot express. Third, existing methods rely on the agent to passively formulate queries to retrieve experiences or skills, whereas PRAXIS proactively surfaces relevant information when the agent touches the associated code, addressing the fundamental problem that agents cannot search for what they do not know they lack.
\section{Threats to Validity}
\label{sec:threats}

Agent-based methods typically involve several hyperparameter settings. Due to the significant cost of LLM APIs, we did not perform an exhaustive hyperparameter search. However, even with settings chosen by intuitive heuristics, PRAXIS consistently achieved the best results across all evaluated configurations, and further tuning would likely yield additional gains. Additionally, using agent workflows to simulate human workflows to extract tacit knowledge means that some gap may remain between the agent's practice process and real human development. Nevertheless, the agent-based approach eliminates manual labor and readily scales performance through large-scale generation. Finally, our evaluation is conducted on KoCo-Bench and AInsteinBench, which collectively cover multiple real-world projects across diverse domains. While these benchmarks may not fully represent the complexity of all possible domain-specific codebases, extending PRAXIS to more domains remains a direction for future work.
\section{Conclusion}
\label{sec:conclusion}

In this paper, we identified tacit knowledge as the fundamental bottleneck in domain code generation and characterized three inherent properties that make it particularly resistant to existing approaches. We presented PRAXIS, a framework that systematically extracts, organizes, and delivers tacit knowledge for domain code generation. PRAXIS acquires tacit knowledge by simulating human development workflows within the target codebase, organizes it on the code dependency graph to preserve cross-component dependencies, and proactively surfaces it during the agent's workflow without requiring the agent to recognize its own knowledge deficit. Experiments demonstrate that PRAXIS outperforms several representative agents and state-of-the-art experience-based and skill-based methods, generalizes across agent frameworks and LLMs, with performance continuing to scale as practice accumulates and online evolution proceeds. As general-purpose code generation capabilities advance rapidly, we believe that the ability to acquire and accumulate domain expertise through practice will become an essential capability for autonomous software engineering agents.

\section*{Acknowledgment}
This research is supported by the National Key R\&D Program under Grant No.~2023YFB4503801, the National Natural Science Foundation of China under Grant Nos.~62192733, 62192730, and 62192731, and the Beijing Major Science and Technology Project under Contract No.~Z251100008425005.
We also acknowledge support from the CIE-Tencent Doctoral Research Incentive Project (Hunyuan Large Model Special Track).

\bibliographystyle{IEEEtran}
\bibliography{ref}

\end{document}